# Bidirectional swapping quantum controlled teleportation based on maximally entangled five-qubit state

ZHA Xin-Wei[*], SONG Hai-Yang Ma Gang-Long

School of Science, Xi'an Institute of Posts and Telecommunications, Xi'an, 710121, P R China

**Abstract** A theoretical scheme for bidirectional swapping quantum controlled teleportation is presented using the entanglement property of maximally entangled five-qubit state, i.e., Alice wants to transmit a entangled state of particle a to Bob and Bob wants to transmit a entangled state of particle b to Alice via the control of the supervisor Charlie.

Since the first creation of quantum teleportation protocol by Bennett [1], research on quantum teleportation has been attracting much attention both in theoretical and experimental aspects in recent years due to its important applications in quantum calculation and quantum communication. Several experimental implementations [2-4] of teleportation have been reported and some schemes of quantum teleportation have also been presented [6-10]. In 2000, Zhou *et al* presented a scheme for controlled quantum teleportation [11].

In this paper, we present a bidirectional swapping quantum controlled teleportation scheme in which a maximally entangled five-qubit state initially shared by the sender (receiver) Alice, Bob and supervisor Charlie functions as a quantum channel.

Recently, Brown *et al.* [12] arrived at a maximally entangled five-qubit state through an extensive numerical optimization procedure. This has the form:

$$|\psi\rangle_{12345} = \frac{1}{2}(|001\rangle|\phi_-\rangle + |010\rangle|\psi_-\rangle + |100\rangle|\phi_+\rangle + |111\rangle|\psi_+\rangle)_{12345} \quad (1)$$

where,

---

[*] E-mail: zhxw@xupt.edu.cn

$$|\psi_\pm\rangle = \frac{1}{\sqrt{2}}(|00\rangle \pm |11\rangle), \text{ and } |\phi_\pm\rangle = \frac{1}{\sqrt{2}}(|01\rangle \pm |10\rangle) \qquad (2)$$

are Bell states.

Substituting formulas (2) into Eq. (1), Eq. (1) can be rewritten as

$$|\psi_5\rangle_{12345} = \frac{1}{2\sqrt{2}}(|00101\rangle - |00110\rangle + |01000\rangle - |01011\rangle \\ + |10001\rangle + |10010\rangle + |11100\rangle + |11111\rangle)_{12345} \qquad (3)$$

If Alice, Bob, and Charlie have particles $12, 34, 5$, the quantum channel can be expressed as:

$$|\psi_5\rangle_{A_1 A_2 B_1 B_2 C} = \frac{1}{2\sqrt{2}}(|00101\rangle - |00110\rangle + |01000\rangle - |01011\rangle \\ + |10001\rangle + |10010\rangle + |11100\rangle + |11111\rangle)_{A_1 A_2 B_1 B_2 C} \qquad (4)$$

Suppose that Alice has particle a in an unknown state:

$$|\chi\rangle_a = (a_0|0\rangle + a_1|1\rangle)_a \qquad (5a)$$

And that Bob has particle b in an unknown state:

$$|\chi\rangle_b = (b_0|0\rangle + b_1|1\rangle)_b \qquad (5b)$$

Alice wants to transmit particle a to Bob and Bob wants to transmit particle b to Alice.

The system state of the seven particles can be expressed as:

$$|\psi\rangle_s = |\chi\rangle_a \otimes |\chi\rangle_b \otimes |\psi_5\rangle_{A_1 A_2 B_1 B_2 C} \qquad (6)$$

In order to realize teleportation, first Alice performs a unitary transformation under the basis $\{|00\rangle, |01\rangle, |10\rangle, |11\rangle\}$, the unitary transformation may take the form

$$U_{A_1 A_2} = \begin{pmatrix} 0 & 1 & 0 & 0 \\ 0 & 0 & 1 & 0 \\ 0 & 0 & 0 & 1 \\ -1 & 0 & 0 & 0 \end{pmatrix} \qquad (7)$$

which will transform Eq. (4) to the result

$$|\psi_5'\rangle_{A_1A_2B_1B_2C} = U_{A_1A_2}|\psi_5\rangle_{A_1A_2B_1B_2C} = \frac{1}{2\sqrt{2}}(-|11101\rangle + |11110\rangle + |00000\rangle - |00011\rangle$$
$$+ |01001\rangle + |01010\rangle + |10100\rangle + |10111\rangle)_{A_1A_2B_1B_2C} \tag{8}$$

Then Alice and Bob have to perform Bell --state measurements on qubit pairs $(A_1, a)$, $(B_2, b)$,

Respectively. Finally, Charlie performs a Von Neumann measurement on his single qubit.

In accordance with the principle of superposition, $|\psi_5'\rangle_{A_1A_2B_1B_2C}$ can be represented in the

following form [13]

$$|\psi\rangle_s = |\chi\rangle_a |\psi\rangle_{A_1A_2B_1B_2C} |\chi\rangle_b$$
$$= \frac{1}{4\sqrt{2}} \sum_{i=1}^{4} \sum_{j=1}^{4} \sum_{k=0}^{1} |\varphi_C^k\rangle |\varphi_{A_1a}^i\rangle \hat{\sigma}_{B_1}^{k,i} |\chi\rangle_{B_1} |\varphi_{bB_2}^j\rangle \hat{\sigma}_{A_2}^{k,j} |\chi\rangle_{A_2} \tag{9}$$

Where $|\chi\rangle_{B_1} = (a_0|0\rangle + a_1|1\rangle)_a$, $|\chi\rangle_{A_2} = (b_0|0\rangle + b_1|1\rangle)_b$ (10)

and $|\varphi_{A_1a}^{j_1}\rangle$, $|\varphi_{B_2b}^{j_2}\rangle$ are Bell states, i.e.

$$|\varphi^1\rangle_{mn} = \frac{1}{\sqrt{2}}(|00\rangle + |11\rangle)_{mn},$$

$$|\varphi^2\rangle_{mn} = \frac{1}{\sqrt{2}}(|00\rangle - |11\rangle)_{mn},$$

$$|\varphi^3\rangle_{mn} = \frac{1}{\sqrt{2}}(|01\rangle + |10\rangle)_{mn}, \quad mn = aA_1, B_2b$$

$$|\varphi^4\rangle_{mn} = \frac{1}{\sqrt{2}}(|01\rangle - |10\rangle)_{mn} \tag{11}$$

.and $|\varphi_c^0\rangle = |0\rangle_c$, $|\varphi_c^1\rangle = |1\rangle_c$.

As we pointed, the operator $\hat{\sigma}_{B_1}^{k,i}$ and $\hat{\sigma}_{A_2}^{k,j}$ are called the transformation operators or "collapse operators". If collapse operators $\hat{\sigma}_{B_1}^{k,i}$ and $\hat{\sigma}_{A_2}^{k,j}$ are unitary operator, According to the outcomes

received, Alice and Bob can determine the state of particles $A_2, B_1$ exactly by the inverse of the

collapse operators $(\hat{\sigma}_{A_2}^{k,j})^{-1}$ and $(\hat{\sigma}_{B_1}^{k,i})^{-1}$. The unknown particle entangled state can be teleported respectively, so teleportation is successfully realized perfectly.

After Alice and Bob two Bell–state measurements and Charlie Von Neumann measurements, the corresponding collapsed state of particle $B_1, A_2$ will be $\dfrac{1}{4\sqrt{2}} \hat{\sigma}_{B_1}^{k,i} |\chi\rangle_{B_1} \hat{\sigma}_{A_2}^{k,j} |\chi\rangle_{A_2}$.

For example, if Alice's measurements result is $|\varphi_{A_1a}^1\rangle$, Bob's measurements result is $|\varphi_{B_2b}^1\rangle$, the corresponding collapsed state of particle $B_1, A_2, C$ will be

$$|\psi^{11}\rangle_{A_2 B_1 C} = \langle \varphi_{A_1a}^1 | \langle \varphi_{B_2b}^1 | \psi \rangle_s$$
$$= \frac{1}{4\sqrt{2}} [a_0 b_0 (|000\rangle + |101\rangle) + a_0 b_1 (|100\rangle - |001\rangle) \qquad (12)$$
$$+ a_1 b_0 (|010\rangle - |111\rangle) + a_1 b_1 (|110\rangle - |011\rangle)]_{A_2 B_1 C}$$

Then, if Charlie's Von Neumann measurement result is $|0\rangle_c$, the collapse operators $\hat{\sigma}_{B_1}^{0,1} = I_{B_1}$ and $\hat{\sigma}_{A_2}^{0,1} = I$, the corresponding collapsed state of particle $B_1, A_2$ will be

$$|\psi^{11,0}\rangle_{A_2 B_1} = {}_C\langle 0 | \psi^{11} \rangle_{A_2 B_1 C} = {}_C\langle 0 | \langle \varphi_{A_1a}^1 | \langle \varphi_{B_2b}^1 | \psi \rangle_s$$
$$= \frac{1}{4\sqrt{2}} (b_0 |0\rangle + b_1 |1\rangle)_{A_2} (a_0 |0\rangle + a_1 |1\rangle)_{B_1} \qquad (13a)$$

Obviously, Alice and Bob do not do anything; the Bidirectional quantum controlled teleportation is successfully realized.

If Charlie's Von Neumann measurement result is $|1\rangle_c$, the collapse operators $\hat{\sigma}_{B_1}^{1,1} = \hat{\sigma}_{B_{1z}}$ and $\hat{\sigma}_{A_2}^{1,1} = i\hat{\sigma}_{A_{2y}}$, the corresponding collapsed state of particle $B_1, A_2$ will be

$$|\psi^{11,1}\rangle_{A_2 B_1} = {}_C\langle 1 | \psi^{11} \rangle_{A_2 B_1 C} = {}_C\langle 1 | \langle \varphi_{A_1a}^1 | \langle \varphi_{B_2b}^1 | \psi \rangle_s$$
$$= \frac{1}{4\sqrt{2}} i\hat{\sigma}_{A_{2y}} (b_0 |0\rangle + b_1 |1\rangle)_{A_2} \hat{\sigma}_{B_{1z}} (a_0 |0\rangle + a_1 |1\rangle)_{B_1} \qquad (13b)$$
$$= \frac{1}{4\sqrt{2}} (b_0 |1\rangle - b_1 |0\rangle)_{A_2} (a_0 |0\rangle - a_1 |1\rangle)_{B_1}$$

Then, Alice and Bob operate unitary transformation $\hat{\sigma}_{A_{2y}}, \hat{\sigma}_{B_{1z}}$ on particle $A_2, B_1$ respectively; the

Bidirectional quantum controlled teleportation is successfully realized.

Similarly, according to the results of Alice and Bob two Bell –state measurements and Charlie Von Neumann measurement, Alice and Bob operate an appropriate unitary transformation; the Bidirectional quantum controlled teleportation is easily realized.

In summary, we have proposed a scheme for Bidirectional quantum controlled teleportation by maximally entangled five-qubit state. Our scheme may be widely useful to the future quantum communication and experiment.

Acknowledgements

This work is supported by the National Natural Science Foundation of China (Grant No. 10902083) and Shaanxi Natural Science Foundation under Contract No. 2009JM1007。;